\documentclass[aps,footinbib,superscriptaddress,groupedaddres,showpacs,twocolumn,prb,longbibliography,bibnotes,showkeys]{revtex4-1}

\usepackage{amssymb,amsbsy,amsmath,mathrsfs}
\usepackage{graphicx}
\usepackage{times}
\usepackage{braket}
\usepackage{stackengine}
\usepackage{verbatim}
\usepackage{hyperref}

\def \titlename{Superlattice induced electron percolation within a single Landau level}
\def \authornames{Nilanjan Roy, Bo Peng, and Bo Yang}
\def \affiliations{Division of Physics and Applied Physics, School of Physical and Mathematical Sciences, Nanyang Technological University, Singapore 637371} 

\begin{document}

\title{Superlattice induced electron percolation within a single Landau level}
\author{Nilanjan Roy}
\affiliation{Division of Physics and Applied Physics, Nanyang Technological University,
Singapore 637371}
\author{Bo Peng}
\affiliation{Division of Physics and Applied Physics, Nanyang Technological University,
Singapore 637371}
\author{Bo Yang}
\email{yang.bo@ntu.edu.sg}
\affiliation{Division of Physics and Applied Physics, Nanyang Technological University,
Singapore 637371}

\begin{abstract}
We investigate the quantum Hall effect in a single Landau level in the presence of a square superlattice of $\delta$-function potentials. The interplay between the superlattice spacing $a_s$ and the magnetic length $\ell_B$ in clean system leads to three interesting characteristic regimes corresponding to $a_s<\ell_B$, $a_s\gg \ell_B$ and the intermediate one where $a_s\sim \ell_B$.   
In the intermediate regime, the continuous magnetic translation symmetry breaks down to discrete lattice symmetry. In contrast, we show that in the other two regimes, the same is hardly broken in the topological band despite the presence of the superlattice.  
In the presence of weak disorder (white-noise) one typically expects a tiny fraction of extended states due to topological protection of the Landau level. Interestingly, we obtain a large fraction of extended states throughout the intermediate regime which maximizes at the special point $a_s=\sqrt{2\pi}\ell_B$. We argue the superlattice induced percolation phenomenon requires both the breaking of the time reversal symmetry and the continuous magnetic translational symmetry. It could have a direct implication on the integer plateau transitions in both continuous quantum Hall systems and the lattice based anomalous quantum Hall effect. 

\end{abstract}

\maketitle

\section{Introduction}
Anderson localisation (AL) is one of the most outstanding phenomena in physics, which states that for time reversal symmetric (TRS) systems all the single particle eigenstates become localized in one and two dimensions in the presence of random disorder~\cite{anderson1958absence}. The first experimental evidence of AL was found in integer quantum Hall effect (IQHE)~\cite{klitzing1980new} where magnetic field breaks TRS and help form flat energy bands, known as the Landau Levels (LL). The impurities or random disorder localise almost all the states in the LL, leading to the vanishing of longitudinal conductance when the filling fraction goes though the localised states. However, every LL is also a topological band with Chern number $C=1$ and hence typically one finds one or very few extended states at the middle of the band that can carry the bulk current giving rise to peak in the longitudinal conductance and a jump in the Hall conductance~\cite{laughlin1981quantized,thouless1982quantized,prange1981quantized,feldman2016observation,ippoliti2018integer,slevin2009critical,huo1992current}. Such localization plays a crucial role for the experimental observation of the quantized Hall plateau in experiments, where the desirable large width of the plateau requires the scarcity of the delocalised states. 

From a percolation theory perspective, delocalisation is equivalent to having an infinite cluster of sites occupied with finite probability that spans across the entire system. Incorporating the classical percolation theory~\cite{saleur1987exact} with the quantum tunneling and interference effect, a random network model,namely the Chalker-Coddington model was proposed~\cite{chalker1988percolation,lee1993quantum}. The model explains the localised to extended state transition in a single LL through a one parameter scaling theory which predicts that the localisation length diverges at the transition with a power-law exponent. The percolation picture and exponent were in agreement with the experimental findings~\cite{koch1991size,hashimoto2008quantum} although there have been a lot of debate later on about making this agreement even more exact.

In the past, it has been shown theoretically~\cite{prange1981quantized} and later on verified experimentally~\cite{feldman2016observation} that a single $\delta$-potential as impurity splits a single localized bound state from a single lowest Landau level (LLL), while the value of Hall conductance remains the same. A superlattice of many $\delta$-potentials has also been recently~\cite{ippoliti2018integer} proposed where the lattice spacing $a_s$ is much larger than the magnetic length $\ell_B$. This shallow super-lattice gives rise to two bands: a topological Chern band ($C=1$) and the other non-topological band consisting of the number of states equal to the number of $\delta$-potentials and concentrating around the energy of single $\delta$-potential. Hence weak disorder localizes the non-toplogical band whereas only very few midband states of Chern band delocalizes. 

\begin{figure}
\centering
\includegraphics[width=1.0\columnwidth,height=4.4cm]{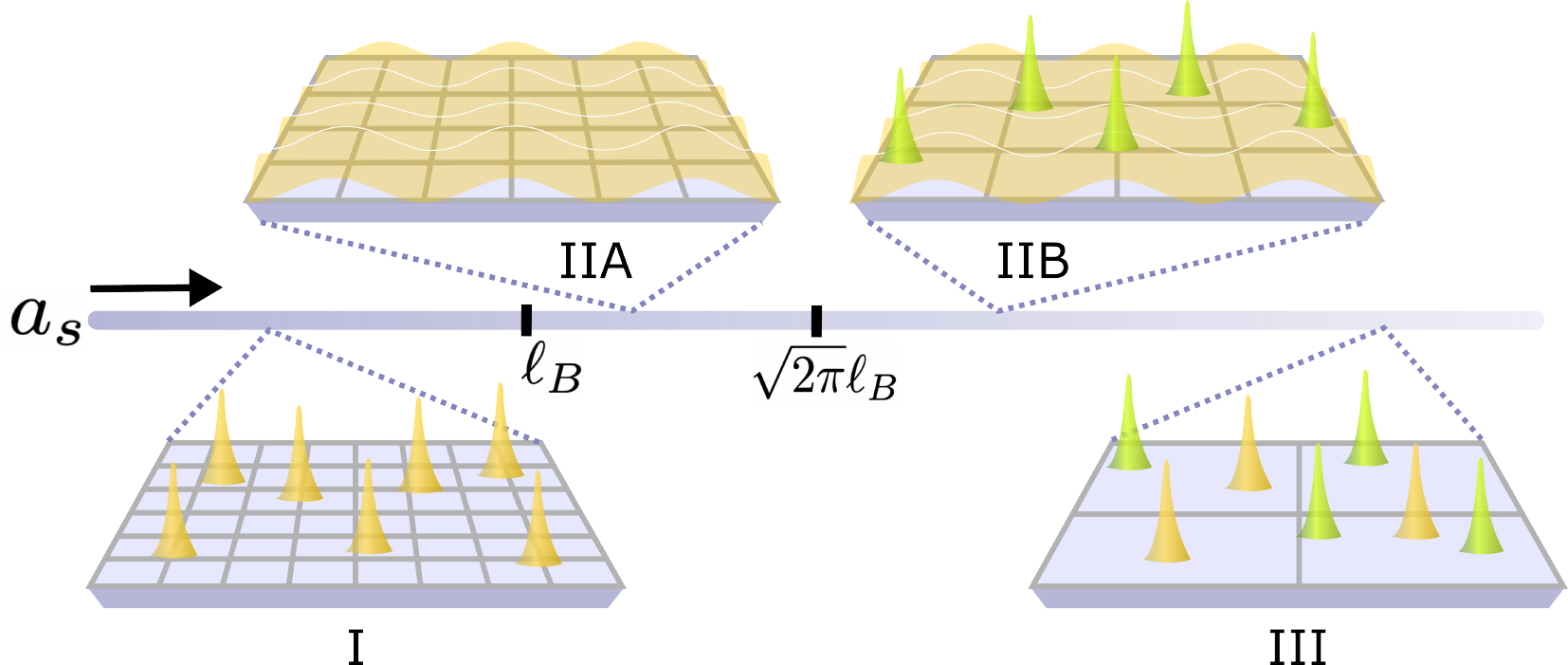}
\caption{{\bf Schematic of the system and main results}: Our system in the LLL has two length scales: the lattice spacing $a_s$ of the superlattice of $\delta$-potentials and constant magnetic length $\ell_B=\sqrt{\hbar/eB}$, where $B$ is the external magnetic field. The schematic shows the essential physics in the regions I, IIA, IIB and III (see caption of Fig.~\ref{clean_system} for definitions) obtained by increasing $a_s$ as shown by the bar in the middle, the left and right ends of which represent the denser and thinner lattices, respectively. In the schematic the grids represent the superlattice of $\delta$-potentials. In regions I and IIA the Bloch states (yellow) are localized and delocalized, respectively. In region IIB, Bloch states are delocalized but coherent states (green) are localized whereas in region III, both the coherent states and Bloch states are localized. }
\label{schematic_regions}
\end{figure}

It is also interesting to consider the limit where $a_s\sim\ell_B$. In such a case the degenerate Landau level evolves to a topological Bloch-band with the bandwidth quickly vanishing as the lattice spacing is reduced. This is because a $\delta$-potential within the LLL localise a coherent state, and a von Neumann lattice (vNL) of coherent states with $a_s=\sqrt{2\pi}\ell_B$ form a complete basis of the LLL~\cite{dana1983adams,ishikawa1995vNL,yang2020vNL}. As $a_s$ decreased below $\sqrt{2\pi}\ell_B$ the number of $\delta$-potentials becomes larger than the number of flux. The lattice of $\delta$-potentials with $a_s\leq\ell_B$ can thus be understood as the vNL Hamiltonian that hardly breaks the translational symmetry within the LLL~\cite{pengbo}, a regime that has not been previously explored; it could be important for superlattice band engineering and topological bands in lattice systems
where the continuous translation symmetry is naturally broken.

In this Letter, we consider localization and percolation of electrons in a \emph{continuous two-dimensional system} subject to a strong out-of-plane magnetic field, and also in the presence of a square superlattice of positive $\delta$-potentials. Since we are interested in the bulk properties, we consider a two dimensional plane with periodic boundary conditions i.e. the torus geometry. Very interestingly, in the intermediate region with $a_s\sim\ell_B$ we find very robust evidence that most of the single particle states are \emph{strongly delocalised in the bulk}, leading to extensive percolation and the absence of the Anderson localisation. This obstruction to the Anderson localisation is due to the reduction of the continuous magnetic translation symmetry (CMTS) to the discrete magnetic translational symmetry (DMTS), where the broken time reversal symmetry and thus the unique properties of the single LL also play an important role. Our main results are schematically presented in Fig.~\ref{schematic_regions}.  

The presence of a large fraction of delocalised states would have significant experimental ramifications in the transport measurement, especially with respect to the width and robustness of the quantum Hall plateau. The interplay between different length scales in topological bands on the hitherto unexplored localisation properties of electrons warrants detailed theoretical and experimental studies both for strongly correlated topological phases, and for topological phases realized in zero magnetic field lattice systems such as the anomalous quantum Hall effect~\cite{haldane1988model,chang2023colloquium}. 
Our analyses make the presence of non-uniform distribution of (large) local Berry curvature distribution responsible for the less robust quantum Hall plateaus obtained in such lattice systems.

\section{The model}
We consider a two-dimensional non-interacting electron gas in a strong perpendicular magnetic field with the following Hamiltonian projected to the LLL.
\begin{eqnarray}
 H = V(\Vec{r}) + \lambda\sum_{i=1}^{N_\delta} \delta(\Vec{r}-\Vec{r}_i),
 \label{ham}
\end{eqnarray}
where the first term stands for disorder potential, which is an uncorrelated white-noise potential with strength $W$ such that $\langle V(\Vec{r})V(\Vec{r^\prime})\rangle=W^2 \delta(\Vec{r}-\Vec{r^\prime})$ as explained later. The second term represents the superlattice potential consisting of $N_\delta$ number of positive delta-potentials of strength $\lambda$ where $N_\delta=L^2$, the number of sites on a square lattice. We work in the weak disorder limit and hence $W\ll\lambda$ so that the disorder can be treated as a small perturbation.

Such two-dimensional system is characterized by two length scales: the magnetic length $\ell_B=\sqrt{\frac{\hbar}{eB}}$ from the external magnetic field and the superlattice spacing $a_s$. The presence of magnetic field gives rise to the magnetic translation operator $\hat{\tau}(\Vec{R})=e^{\Vec{R}.(\Vec{\nabla} - i\Vec{A}/\ell_B^2) - i\hat{z}.(\Vec{R} \times \Vec{r})/\ell_B^2}$ where $\Vec{r}$ is the co-ordinates of electron with the Landau gauge $\Vec{A}=-By\hat{x}$.  With a total of $N_\phi$ magnetic fluxes, the $j\textsuperscript{th}$ wavefunction of the LLL basis on torus is given by~\cite{lee_halperin1983,haldane1985torus,hermanns2008torus},
\begin{eqnarray}
\psi_j(\Vec{r}) = \frac{1}{\sqrt{L_s \ell_B\sqrt{\pi}}} \sum_{k=-\infty}^{\infty} e^{ -i\frac{(\chi_j + kL_s)x}{\ell_B^2} - \frac{(y-\chi_j-kL_s)^2}{2 \ell_B^2}},
\label{wfn2}
\end{eqnarray}
where $\chi_j=2\pi \ell_B^2 j/L_s$ and $0\leq x,y <L_s$, where $L_s=a_sL$. Periodicity implies $L_s^2=2\pi \ell_B^2 N_\phi$ which further implies $\psi_{j+N_\phi}=\psi_j$. Therefore, there are $N_\phi$ linearly independent basis states and hence $j=\{0,1,2,..,N_{\phi}-1\}$.
For numerical purpose, $V(\Vec{r})=\sum_{i=1}^{N_\epsilon} W_i \delta(\Vec{r}-\Vec{r}_i)$ where the locations of $N_\epsilon$ number of delta functions are chosen randomly from a uniform distribution and there are equal number of delta
functions with positive $(W_i=W)$ and negative $(W_i=-W)$ magnitudes respectively. In the limit $N_\epsilon\gg N_\phi$, this distribution mimics an uncorrelated white noise potential with mean $\langle V(\Vec{r})\rangle=0$ and correlation $\langle V(\Vec{r})V(\Vec{r^\prime})\rangle=W^2 \delta(\Vec{r}-\Vec{r^\prime})$ and also keeps the density of states symmetric with respect to the center of the Landau level without the superlattice~\cite{huckestein1995scaling}. 
\begin{figure}
\centering
\stackon{\includegraphics[width=0.98\columnwidth,height=4.8cm]{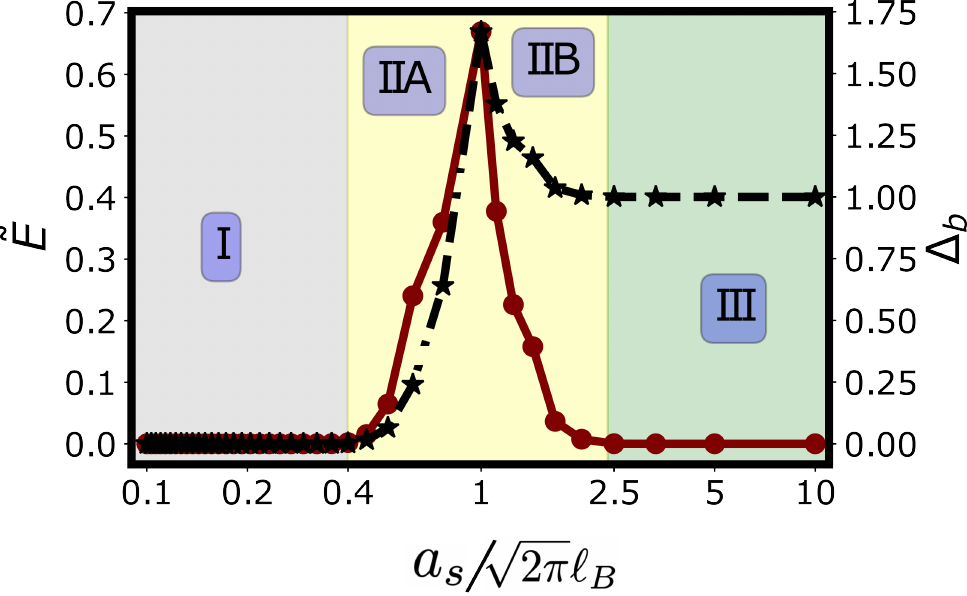}}{(a)}
\stackon{\includegraphics[width=0.495\columnwidth,height=3.7cm]{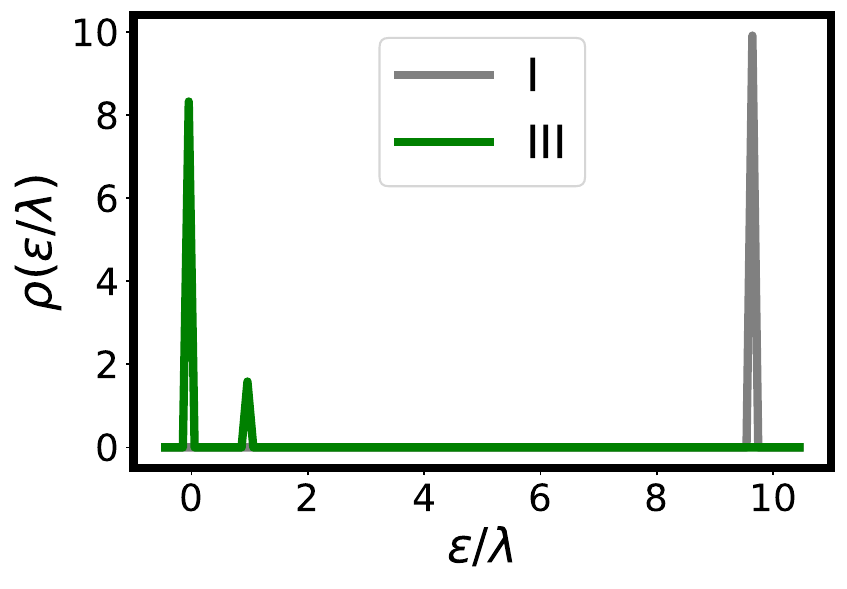}}{(b)}
\stackon{\includegraphics[width=0.495\columnwidth,height=3.7cm]{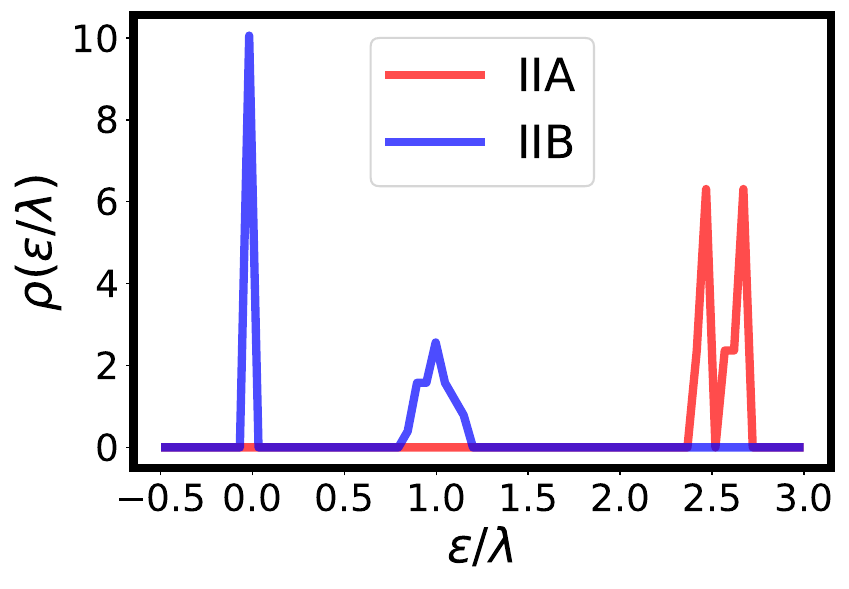}}{(c)}
\caption{
(a) {\bf Effect of lattice on clean system}: $\Tilde{E}$ (solid) as defined in Eq.~\ref{E_tilde} and $\Delta_b$ (dashed) are plotted as a function of lattice spacing $a_s$ (in units of $\sqrt{2\pi}\ell_B$). 
We identify three regions: I. $a_s<\ell_B$, II. $\ell_B<a_s<6 \ell_B$ and III. $a_s>6 \ell_B$. The intermediate region is further divided into two sub-regions: IIA. $a_s<\sqrt{2\pi}\ell_B$ and IIB. $a_s>\sqrt{2\pi}\ell_B$ where both the plots maximize at the vNL point ($a_s=\sqrt{2\pi}\ell_B$) indicating maximum influence of discrete lattice symmetry. 
{\bf Density of states in clean system with lattice}: (b) Density of states $\rho(\epsilon/\lambda)$ of energies $\epsilon$ re-scaled by the strength of lattice potential $\lambda$ in regions I and III, respectively. (c) The same plots but in regions IIA and IIB, respectively. For all the plots $N_\phi=100$ and $\lambda=10$.
}
\label{clean_system}
\end{figure}

\section{Symmetry in the clean limit}
Every $\delta$-potential on the lattice takes away one state from the degenerate band of zero-energy states~\cite{haldane1988localization} within the LL. Hence the non-zero energy state now appears as a part of the Bloch band far away from zero-energy states (since $\lambda\gg0$)~\cite{ippoliti2018integer} .
In Fig.~\ref{clean_system}(a), we show the energy-bandwidth $\Delta_b$ as a function of lattice spacing $a_s$ (in units of $\ell_B$). From the behavior of $\Delta_b$, one can divide the parameter space (in this case lattice spacing) mainly into three regions: I. $a_s<\ell_B$, II. $a_s\sim \ell_B$ and III. $a_s\gg\ell_B$. 
Regions I and III being two extreme scenarios, can be understood from physical arguments. In region I, the lattice-spacing $a_s<\ell_B$ and there are many $\delta$-potentials $(N_\delta\gg N_\phi)$ whereas $\ell_B$ effectively gives the spatial resolution within a single LL. Hence, the superlattice \emph{hardly breaks} the CMTS within a single LL, even though it explicitly breaks translational symmetry in real space (the full two-dimensional Hilbert space). Hence, the density of states (DOS) looks like that of a flat-band at a non-zero (far away from zero) energy in Fig.~\ref{clean_system}(b). In region III, the number of $\delta$-potentials of the superlattice is very small $(N_\delta\ll N_\phi)$ such that $a_s\gg \ell_B$. The non-zero energy states localised by the $\delta$-potentials have vanishing overlaps between neighbouring sites, leading to exponentially small bandwidth and an essentially flat Bloch band (see Fig.~\ref{clean_system}(a)). The corresponding DOS shows a tiny peak at the non-zero energy (of a single $\delta$-potential) along with a large zero-energy peak representing the original flat-band shown in Fig.~2(b). Hence, in both the regions I and III there is effective CMTS for electrons in a single LL: there are essentially no dispersive single particle states.

Region II on the other hand is qualitatively different. Here $a_s\sim \ell_B$ thus almost every electronic state is localised by a $\delta$-potential, and the overlap between the neighbouring coherent states is substantial, leading to a dispersive Bloch band and only the DMTS. We further divide this part into two sub-regions: IIA where $N_\delta>N_\phi$ leads to $a_s<\sqrt{2\pi}\ell_B$ and IIB where $N_\delta<N_\phi$ leads to $a_s>\sqrt{2\pi}\ell_B$. In region IIA, the zero-energy band is absent and one is only left with the Bloch-band of large bandwidth, which increases with $a_s$ as shown in Fig.~\ref{clean_system}(a). In this region the DOS is that of a dispersive band far away from zero-energy shown in Fig.~\ref{clean_system}(c). In fact, in the limit $N_\delta=N_\phi$ or at $a_s=\sqrt{2\pi}\ell_B$, the bandwidth reaches the maximum. In region IIB, $N_\delta<N_\phi$ and hence one still has the zero-energy flat-band along with the non-zero energy states of Bloch band. Hence, the corresponding DOS has a peak at zero energy and a distribution around non-zero energy (of a single $\delta$-potential) as shown in Fig.~\ref{clean_system}(c). Now as $N_\delta$ ($a_s$) is decreased (increased) the width of Bloch band becomes narrower leading to bandwidth getting smaller and reaching the limit of region III.
Hence, the DOS shown in Fig.~\ref{clean_system}(b) and Fig.~\ref{clean_system}(c) for regions I, III and regions IIA, IIB, respectively provide a complementary picture giving a more clear visual insight into spectrum of clean system. The DOS and bandwidth of vNL corresponding to the Hamiltonian at $a_s=\sqrt{2\pi}\ell_B$ are discussed in more details in the supplementary materials~\cite{suppl}.

To quantitatively capture the effect of the lattice potential we define the following quantity:
\begin{eqnarray}
 \tilde E=\text{min}(\Delta_b,|E_\delta-\Delta_b|)
 \label{E_tilde}
 \end{eqnarray} 
 where $E_\delta$ (here $\lambda=10$) is the energy of a coherent state from a single $\delta$-potential. In region I, $\tilde E$ vanishes since $\Delta_b\approx0$ whereas in region III, again this quantity almost vanishes since $\Delta_b\approx E_\delta$ in this region indicating that the region III is essentially dominated by single $\delta$-potential physics rather than physics of a lattice of many $\delta$-potentials. Now, in region II, due to effect of lattice-induced DMTS, the Bloch band is widened enough and the quantity as shown in Fig.~\ref{clean_system}(a) reflects the deviation from the single $\delta$-potential physics towards the physics of a lattice of many $\delta$-potentials, which reaches maximum at the vNL point, where Bloch band has the maximum width as discussed earlier. Thus $\tilde E$ is a good measure of the amount of the breaking of the CMTS in the system.  

\section{Lattice induced percolation} 
We now look at the effect of weak disorder, which is the Gaussian white-noise as explained previously, on different regions as described above. 
The presence of random disorder leads to broadening of the otherwise flat LLL affecting the localisation properties of the single particle states. Since the Hamiltonian matrix obtained in such a case contains randomness in its elements, we resort to measures from the random matrix theory to analyse the localisation properties. Namely, we study the energy-resolved level-spacing ratio calculated using the consecutive energy-level spacing $s_k = \epsilon_{k+1} - \epsilon_k$ within an energy-window when $\epsilon_k$'s are arranged in the ascending order. Important information about localisation and percolation of the system can be extracted by studying the probability distribution $P(s)$ where $s$ is normalised such that mean level-spacing $\langle s\rangle=1$. In the localised phase, degenerate states can coexist which leads to the appearance of the Poisson distribution $P(s)=e^{-s}$ with a macroscopic value at $s=0$. On the contrary, in the delocalised phase degenerate states are strictly not allowed that leads to distributions $P(s)=A_\beta s^\beta e^{-B_{\beta} s^2}$ which are characterised by level repulsion, i.e. zero value at $s=0$. Here $\beta=1, 2$ and $4$ correspond to Gaussian orthogonal ensemble (GOE) in TRS abiding systems, Gaussian unitary ensemble (GUE) in TRS broken systems and Gaussian symplectic ensemble (GSE) where TRS is preserved but spin rotation symmetry is broken~\cite{mehta2004}. $A_\beta$ and $B_\beta$ are constants required to normalise the respective distribution functions~. 

Presence of TRS is an essential ingredient to obtain Anderson localization in two dimension where all the single particle eigenstates are localised.
However, our system is TRS broken due to the presence of magnetic field. Hence, we expect to obtain a delocalised region in the energy spectrum that corresponds to GUE ensemble, as discussed above. Instead of plotting the whole distribution $P(s)$, one can alternatively calculate the levelspacing ratio $r$, a modern and simpler quantity which does not require spectrum unfolding and gives a single value. The levelspacing ratio is given by~\cite{pal2010mbl},
$r = \langle\langle \frac{\min(s_k,s_{k+1})}{\max(s_k,s_{k+1})} \rangle\rangle$,
where the inner curly braces represent the average over the chosen bin of the energy spectrum and the outer braces stand for the average over the disorder realisations. For the Poisson distribution (localisation) $r=0.386$ whereas $r=0.599$ for the GUE ensemble (delocalisation)~\cite{atas2013,alet2022,pal2010mbl}. It shows intermediate values for the intermediate phases which are neither delocalised nor localised, e.g. the nonergodic extended phases. We calculate the energy-resolved $r(\epsilon)$ to detect different regions in the single particle spectrum.

\begin{figure}
\centering
\stackon{\includegraphics[width=0.495\columnwidth,height=3.4cm]{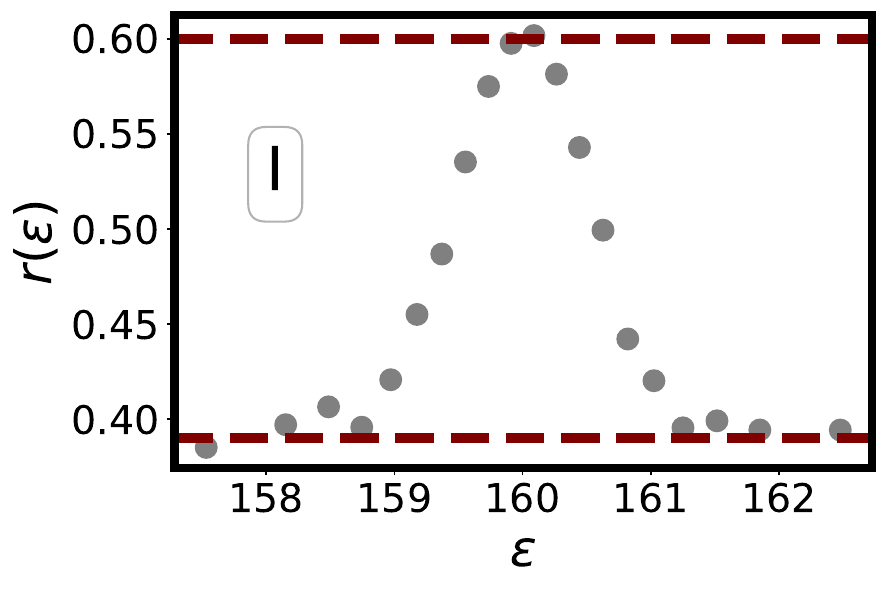}}{(a)}
\stackon{\includegraphics[width=0.495\columnwidth,height=3.4cm]{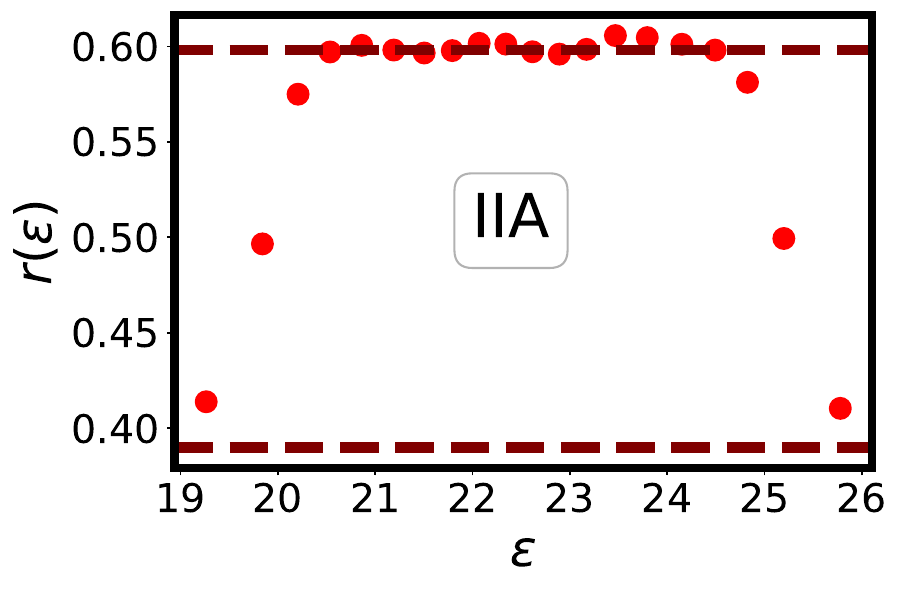}}{(b)}
\stackon{\includegraphics[width=0.495\columnwidth,height=3.4cm]{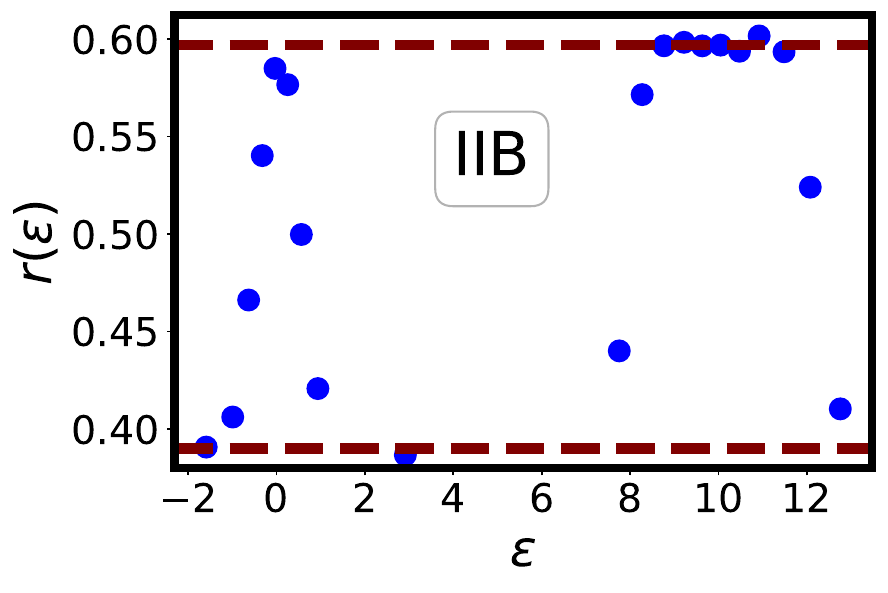}}{(c)}
\stackon{\includegraphics[width=0.495\columnwidth,height=3.4cm]{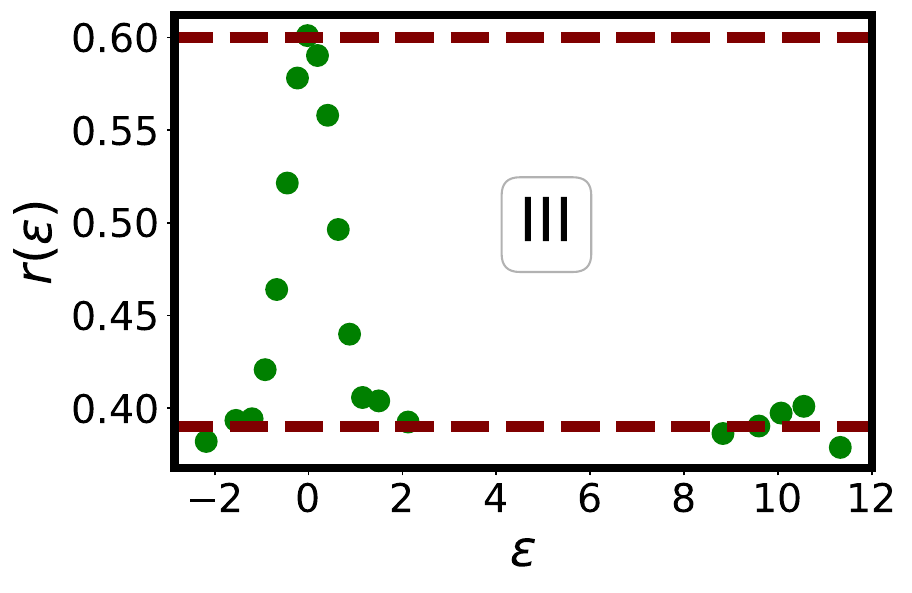}}{(d)}
\caption{ {\bf Energy-resolved levelspacing ratio $r(\epsilon)$ in different regions}: (a-d) $r(\epsilon)$ as a function of single-particle energy $\epsilon$ in regions I, IIA, IIB and III, respectively. 
}
\label{lratio_weak}
\end{figure}

We choose $W=0.2$ and $\lambda=10$ for all numerical calculations in the weak-disorder limit. In Fig.~\ref{lratio_weak}(a-d), we show $r(\epsilon)$ as a function of single-particle energy $\epsilon$ in the regions I, IIA, IIB and III, respectively. In Fig.~\ref{lratio_weak}(a), we see that although most of the states in the Bloch band are now localised with $r\approx0.4$, the mid-band states are delocalised that corresponds to $r\approx0.6$. This is expected in TRS broken systems and consistent with a related previous study~\cite{huo1992current}. Similarly, Fig.~\ref{lratio_weak}(d) shows that very few midband-states in otherwise zero-energy flat-band are delocalised whereas all other states including the ones in otherwise Bloch band tend to get localised. The scenario in region III is expectedly similar to that in region I. 

\begin{figure}
\centering
\stackon{\includegraphics[width=0.495\columnwidth,height=3.7cm]{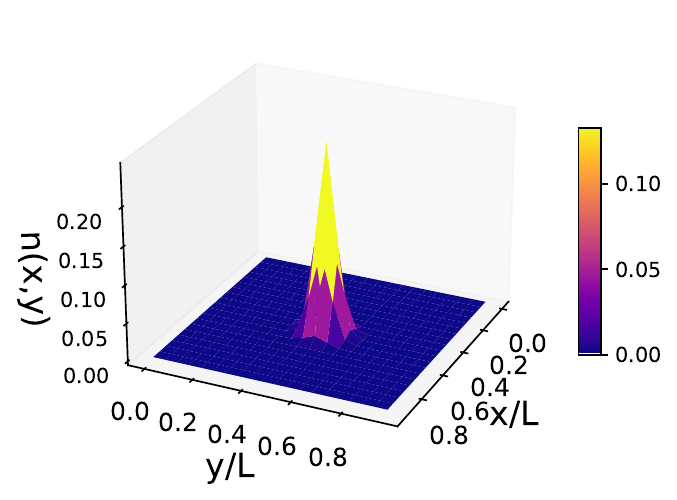}}{(a)}
\stackon{\includegraphics[width=0.495\columnwidth,height=3.7cm]{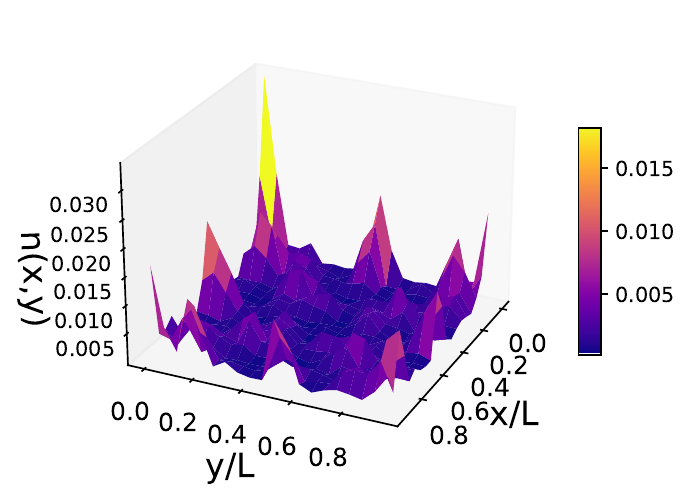}}{(b)}
\stackon{\includegraphics[width=0.495\columnwidth,height=3.5cm]{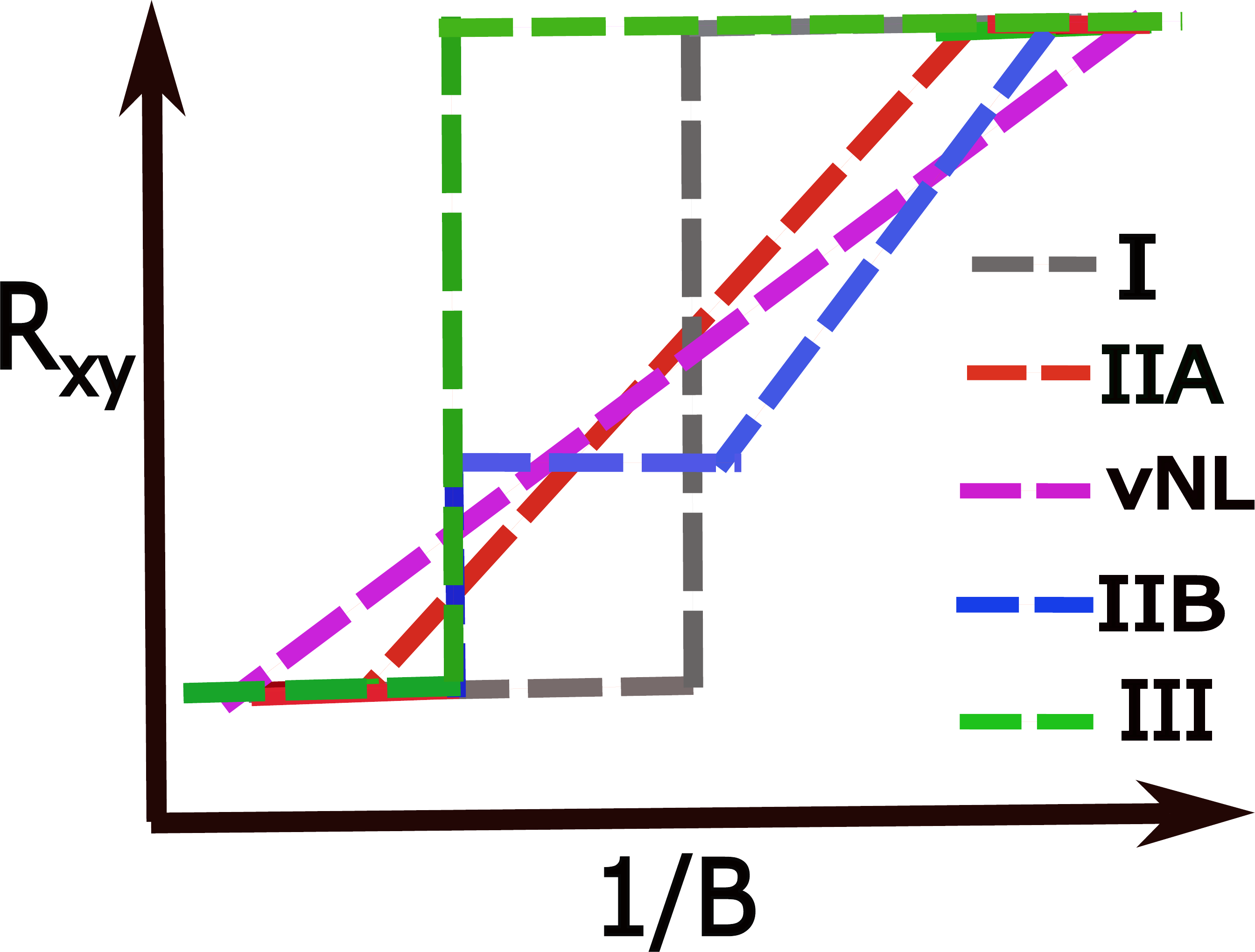}}{(c)}
\stackon{\includegraphics[width=0.495\columnwidth,height=3.6cm]{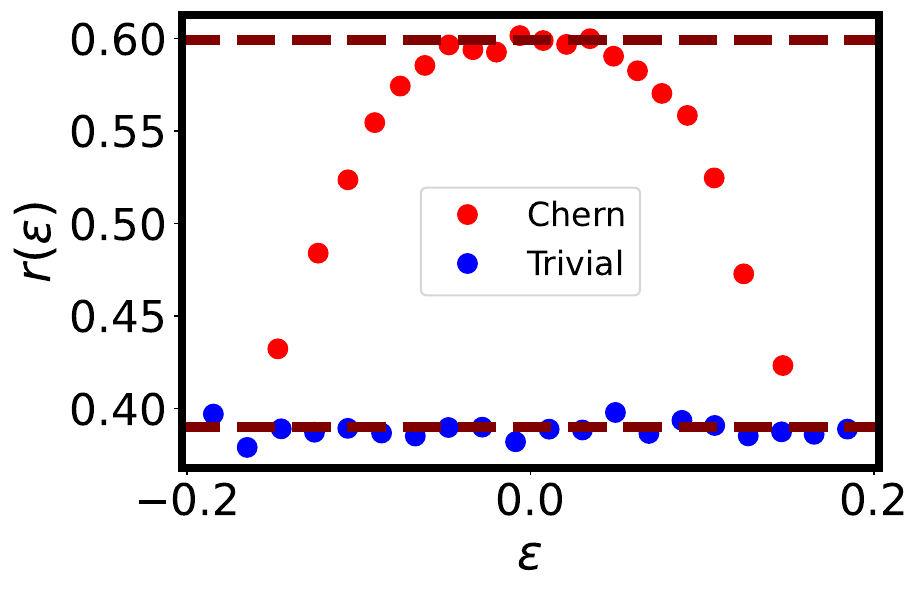}}{(d)}
\caption{ (a-b) {\bf Percolation picture}: Density profile $n(x,y)$ of an initially Gaussian wavepacket in LLL in clean system in the long-time limit in region I (III) and region II, respectively.  (c) {\bf Schematic of expected outcome in experiment}: The expected outcome in the integer quantum Hall experiment as the lattice-spacing $a_s$ is changed such that the system is in regions I, IIA, IIB and III, respectively and also when it is at the vNL limit. 
(d) {\bf Effect of disorder on Chern insulator}: Energy-resolved levelspacing ratio $r(\epsilon)$ as a function of $\epsilon$ in the Chern band and trivial band, respectively in the flat band limit of the weakly disordered Haldane model. 
}
\label{perc_exp}
\end{figure}

The non-trivial scenario happens in regions IIA and IIB, where only DMTS is present. In region IIA, as shown in Fig.~\ref{lratio_weak}(b), there is a large fraction of Bloch band states that are still delocalised, even in the presence of weak disorder whereas very few states at the band-edges tend to localise. In region IIB, both the zero-energy states and Bloch-band states are present. On introducing weak disorder in region IIB, as shown in Fig.~\ref{lratio_weak}(c), while the few the mid-band states near zero energy are delocalised as expected, there is also a large fraction of states in the Bloch band are all delocalised.  An extreme case of the vNL is separately shown in the supplementary material, where almost all the states are delocalised and very robust with respect to finite-size effect~\cite{suppl}. We also provide a dynamical perspective in Fig.~\ref{perc_exp}(a-b) which is the time evolution of an initially Gaussian wavefunction representing a single $\delta$-potential in LLL. As Fig.~\ref{perc_exp}(a) implies that in the long-time limit the wavefunction hardly shifts from a Gaussian in region I or III due to scarcity of delocalised eigenstates whereas in region II, shown in Fig.~\ref{perc_exp}(b), it quickly percolates across the whole system due to dominance of delocalised states in the spectrum.  

We attribute the interesting delocalisation phenomenon in region II to the DMTS imposed onto the system by the dominant superlattice potential strength (than disorder strength). In absence of the DMTS, which can be achieved by either moving to region I or III. or through increasing disorder strength $W$, or by randomizing the positions of $\delta$-potentials instead of a lattice, this enhanced percolation (delocalisation) effect disappears~\cite{suppl}. All of these really stresses on the importance of having an effective DMTS as reason behind the appearance of the large fraction of delocalised states, in a system with \emph{broken TRS}.

\section{Experimental ramifications}
Our results can have direct implications on integer quantum Hall experiments which are mainly explained with the non-interacting single particle physics. The larger fraction of localised states contributes to wider plateaus, while a small fraction of delocalised states leads to steeper plateau-to-plateau transitions and non-vanishing longitudinal resistance. In the presence of a superlattice of $\delta$-potentials with $a_s\sim\ell_B$, the curious absence of the Anderson localization can lead to significant shrinking of the plateaus width and even difficulties in measuring the quantized Hall conductivity (see Fig.~\ref{perc_exp}(c)). We expect that with the recent progress~\cite{hensgens2018capacitance,forsythe2018band,vasilkova2024carrier} with super-lattice potential in two dimensions could lead to the testing of our predictions in the experiments. 
Moreover, it would be interesting to probe the superlattice-induced variation in the sharpness of plateau transitions by measuring the magneto-conductance fluctuations which have been claimed to be multifractal in nature for sharper transitions in the recent four-probe experiments~\cite{amin2022multifractal,barbosa2022turbulence}.

We also expect the same physics of bulk percolation to persist for the fractional quantum Hall (FQH) phases in the presence of the superlattice potential. The interplay between the superlattice constant and the characteristic length scales of the interacting topological phases is fundamental. For the Abelian FQH phases such as the Laughlin phase, such characteristic length scale can be intuitively understood in the composite fermion~\cite{pu2017composite,jain2007composite} picture with the rescaled magnetic field (by flux attachment), leading to a greater $\ell_B^*$ as compared to $\ell_B$. The related subtleties and the characteristic length scales for the non-Abelian topological phases will be discussed in~\cite{pengbo} and they warrant further detailed studies. 

More importantly for lattice based Chern insulators (CI), where a topological Chern band can form at zero magnetic field and local Berry curvature can be large (on the order of $\sim 100 T$)~\cite{wang2006ab,wimmer2017experimental,shin2019unraveling}, periodic potentials from the underlying lattice is naturally present and can no longer be ignored like the cases in quantum Hall systems (where $\ell_B$ is much larger than the lattice constant of the crystal). Thus even without an additional superlattice, there are many relevant length scales in a CI and their effects on the electron localization and percolations in a Chern band are not very well understood. 
In the presence of weak disorder, the Chern band  shows a fraction of extended states (with $r \approx 0.6$), whereas all states in the trivial band become localized (with $r\approx0.39$). We take the flat-band limit (hence ignore kinetic energy) in our calculations as shown in Fig.~\ref{perc_exp}(d). In the Chern band $r$-value indicates that there is a larger fraction of extended states~\cite{suppl}, unlike the results obtained for LLL in our work where a smaller fraction of extended states are found. This can be attributed to the presence of distribution of (large) Berry curvature in the Chern band of CIs.
How will the interplay between these length scales affect the intrinsic robustness of the quantized Hall plateau (both for integer and fractional CIs), and if we can use superlattice engineering~\cite{wang2015evidence,dean2013hofstadter,park2009landau,cai2023signatures}  to enhance the robustness of the plateau from transport measurement~\cite{regnault2011FCI}, can be very exciting and will be discussed elsewhere\footnote{The precision of quantum Hall plateau (ideally $R_{xy}=h/{\nu e^2}$ for $\nu\in$ integers) in CIs is one of the main concerns. Typically this accuracy is $\mathcal{O}(10^{-3})$ in CIs~\cite{chang2015high} whereas for IQHE it is so highly precise up to $\mathcal{O}(10^{-9})$~\cite{schopfer2007testing} that it is being used as the universal standard for resistance metrology. Typically more dirt in the CI samples is made responsible for this than that in the IQHE counterparts. However, efforts have been made recently on CIs with improved precision toward the magnetic-field-independent resistance metrology but they are still not close to the current universal standard~\cite{chang2023colloquium}. }.
Moreover, what happens in presence of correlated disorder can be very interesting, as it adds another length scale to the topological flat band. Our preliminary results~\footnote{We consider Gaussian-correlated disorder potential $V(\Vec{r})$ such that $\langle V(\Vec{r})V(\Vec{r^{\prime}})\rangle = \frac{W^2}{\sigma^2}e^{-|\Vec{r}-\Vec{r^{\prime}}|^2/2\sigma^2}$. Here $W$ is disorder strength and $\sigma$ plays the role of correlation length ($\sigma=0$ for uncorrelated disorder). Our preliminary calculations suggest that the results presented in the main text, for uncorrelated disorder, remain intact as long as $\sigma<\ell_B$. The results start to differ when $\sigma>\ell_B$: the fraction of extended states initially decreases for $\sigma<a_s$ whereas the same fraction starts increasing for $\sigma>a_s$, where $\ell_B$ and $a_s$ are the magnetic length and superlattice spacing, respectively. These will be explored in details in the future.}
suggest that the competition between the correlation length, magnetic length and superlattice constant can potentially lead to even more intricate physics and we will explore it in details in the near future.

\section*{Acknowledgments}
We would like to acknowledge useful discussions with H. Goldman and P. Kumar, and help from Yuzhu Wang for the figures. This work is supported by the NTU grant for the National Research Foundation, Singapore under the NRF fellowship award (NRF-NRFF12-2020-005), and Singapore Ministry of Education (MOE) Academic Research Fund Tier 3 Grant (No. MOE-MOET32023-0003) Quantum Geometric Advantage.

\bibliography{refs}

\def\makeSM{1}
\ifdefined\makeSM
\clearpage
\newpage

\widetext

\appendix
\renewcommand{\appendixname}{}
\renewcommand{\thesection}{{S\arabic{section}}.~}
\renewcommand{\theequation}{\thesection.\arabic{equation}}
\renewcommand{\thefigure}{S\arabic{figure}}

\setcounter{section}{1}
\setcounter{page}{1}
\setcounter{figure}{0}
\setcounter{equation}{0}

\begin{center}
{\bf Supplementary material for ``\titlename" }\\
{\authornames}\\
{\it \affiliations}
\end{center}
\fi 

\section{~\thesection{Density of states and localization behavior of the von Neumann lattice}}
\begin{figure*}[h]
\centering
\stackon{\includegraphics[width=0.4\columnwidth,height=5.3cm]{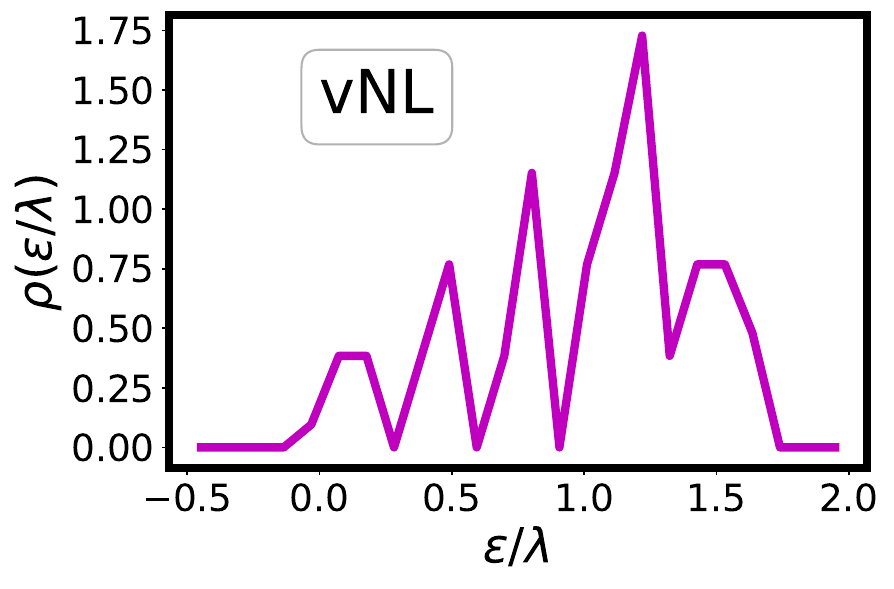}}{(a)}
\stackon{\includegraphics[width=0.4\columnwidth,height=5.3cm]{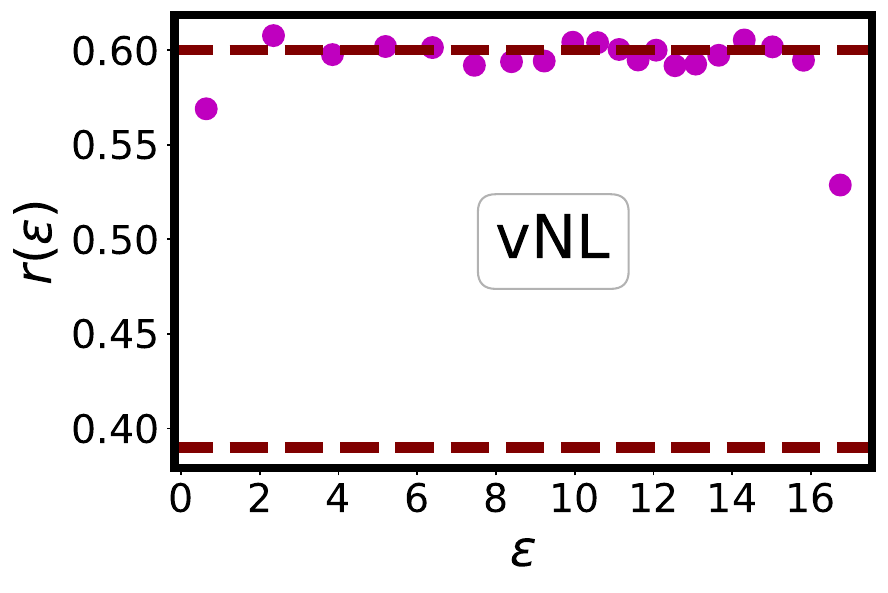}}{(b)}
\caption{ {\bf Results for the von Neumann lattice}: (a) Density of states $\rho(\epsilon/\lambda)$ of energies $\epsilon$ re-scaled by the strength of lattice potential $\lambda$ at vNL limit. $N_\phi=100$ and $\lambda=10$ for this plot. (b) Energy-resolved levelspacing ratio $r(\epsilon)$ as a function of energy $\epsilon$ at the same limit. $N_\phi=1600$, $W=0.2$ and $\lambda=10$ for this plot.}
\label{dos_vNL}
\end{figure*}

Here we discuss the density of states (DOS) of the clean system and localization behavior of the weakly disorder system in the presence of the von Neumann lattice (vNL), i.e. when the number of $\delta$ potentials is exactly equal to the integer number of flux.
In Fig.~\ref{dos_vNL}(a) we show the DOS $\rho(\epsilon/\lambda)$, with re-scaled energies $\epsilon/\lambda$, of the clean $(W=0)$ system at the vNL limit. 
At the vNL point $N_\delta=N_\phi$. Hence the original flat-band at zero energy is just destroyed and the broadening of Bloch-band reaches the maximum, which is evident in Fig.~\ref{dos_vNL}(a) as one compares this with Fig.~2 of the main text. Now as the weak disorder is turned on at this limit almost all states get delocalized as $r\approx0.6$ throughout the spectrum which is shown in Fig.~\ref{dos_vNL}(b). This also shows that the fraction of delocalized states maximizes at the vNL limit as one compares it with other regions which are shown in Fig.~3 in the main text. 

\setcounter{section}{2}
\section{~\thesection{The weak disorder case}}
Here we discuss the dependence of our results, reported in the main text, on the system sizes and the density of disorder potentials. 
\begin{figure}[h]
\centering
\stackon{\includegraphics[width=0.33\columnwidth,height=4.4cm]{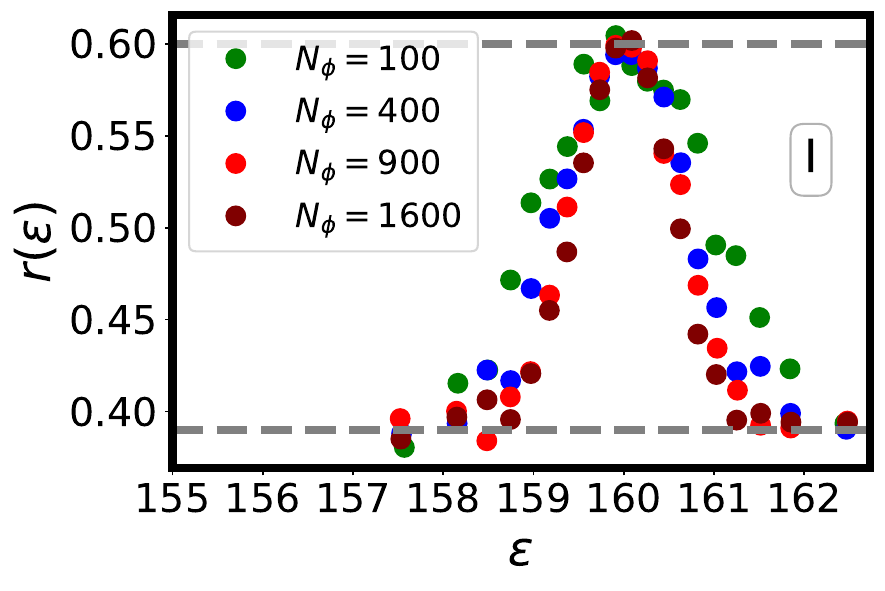}}{(a)}
\stackon{\includegraphics[width=0.33\columnwidth,height=4.4cm]{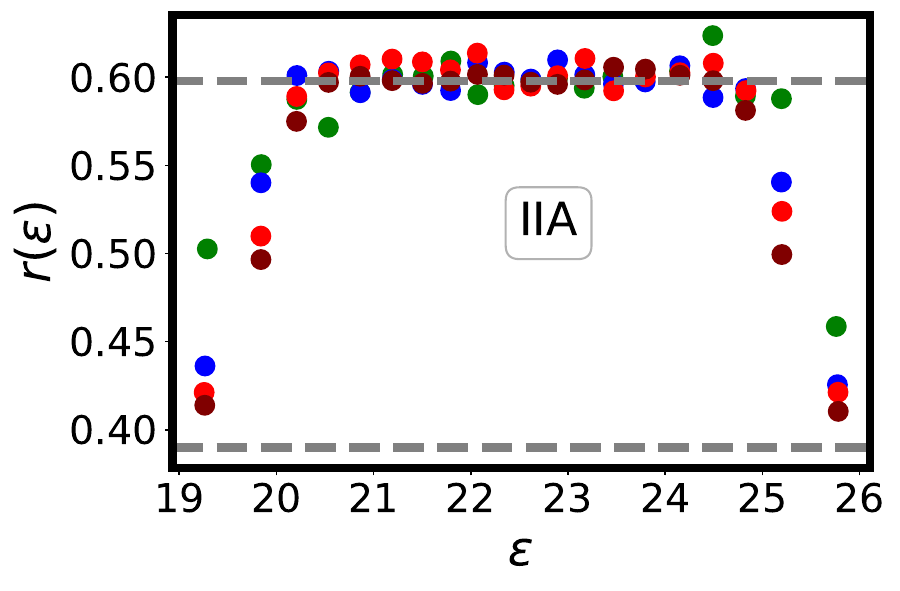}}{(b)}
\stackon{\includegraphics[width=0.33\columnwidth,height=4.4cm]{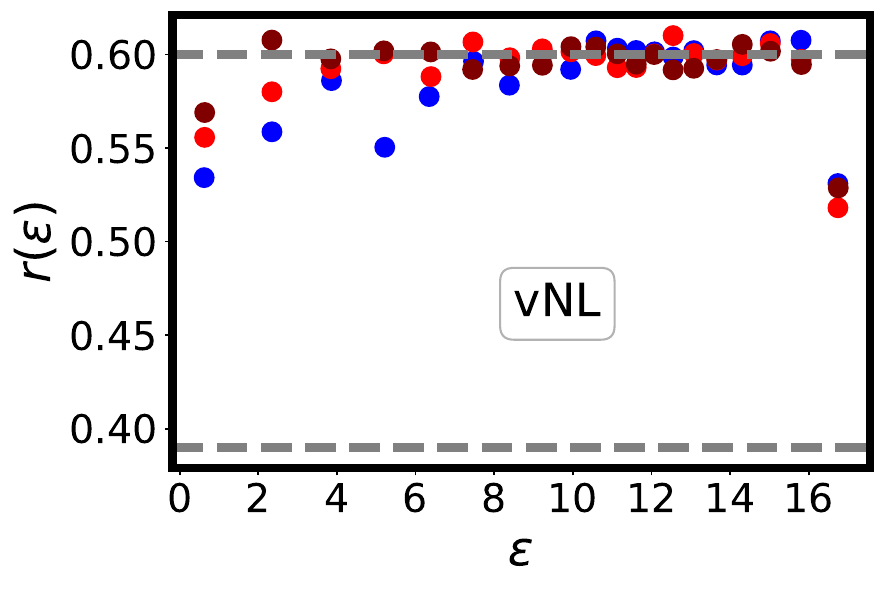}}{(c)}
\stackon{\includegraphics[width=0.33\columnwidth,height=4.4cm]{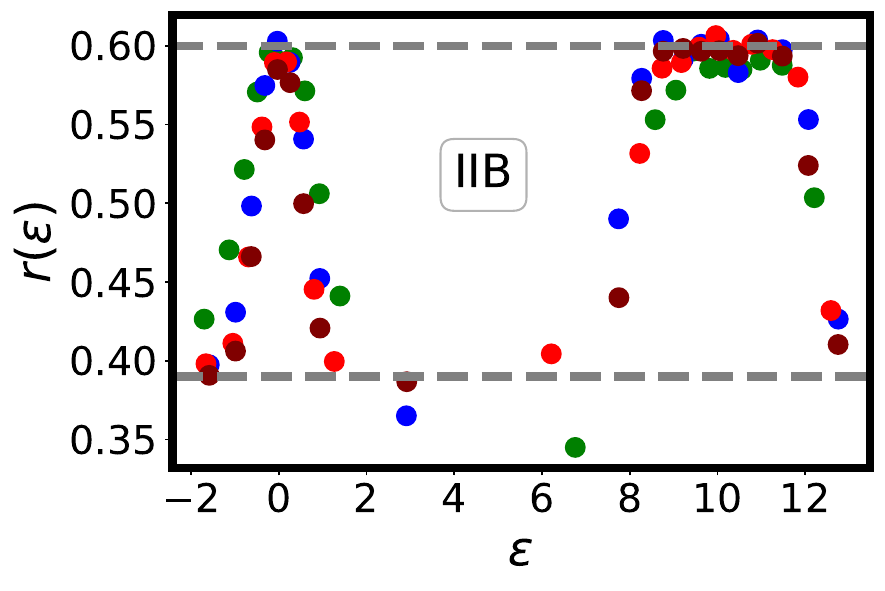}}{(d)}
\stackon{\includegraphics[width=0.33\columnwidth,height=4.4cm]{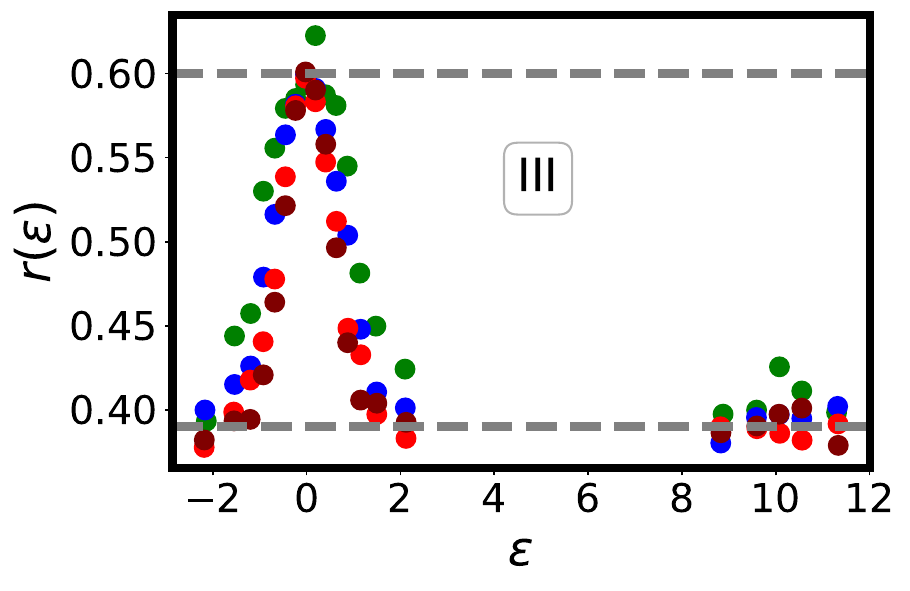}}{(e)}
\caption{ {\bf System-size dependence of energy-resolved levelspacing ratio}: (a-e) $r(\epsilon)$ as a function of $\epsilon$ for increasing system sizes $N_\phi=100,400,900,1600$ in the regions I,IIA, at vNL pint, regions IIB and III, respectively. Here the no. of disorder potentials $N_\epsilon=40 N_\phi$ for all the plots.}
\label{lratio_sysdep}
\end{figure}
In Fig.~\ref{lratio_sysdep}(a-e), we show system-size dependence of our results in regions I, IIA, at vNL point, regions IIB and III, respectively. We find that our result is quite robust with respect to finite size effect and in fact, the same becomes more clear as system size is increased. In Fig.~\ref{lratio_sysdep} in all the plots one has equal number of data points ($20$ points). The no. of states in a single bin is kept proportional to system size $N_\phi$ to obtain a single data point such that the data can be compared for different system sizes as it is increased, especially in region II. For example, in Fig.~\ref{lratio_sysdep}(b) of region IIA, around $70\%$ states are delocalised whereas in Fig.~\ref{lratio_sysdep}(d) of region IIB, around $40\%$ states are delocsalised. At the vNL point, shown in Fig.~\ref{lratio_sysdep}(c) almost $90\%$ or even more states tend to delocalize.  
We find similar results as we now increase the density of disorder potentials for a constant system size, as shown in Fig.~\ref{lratio_disorder}(a-e) for different regions of lattice-spacing including the vNL point. Since, adding more and more disorder potentials helps in approaching the white-noise limit, the plots approach the results claimed in the main text.  

\begin{figure}[h!]
\centering
\stackon{\includegraphics[width=0.33\columnwidth,height=4.4cm]{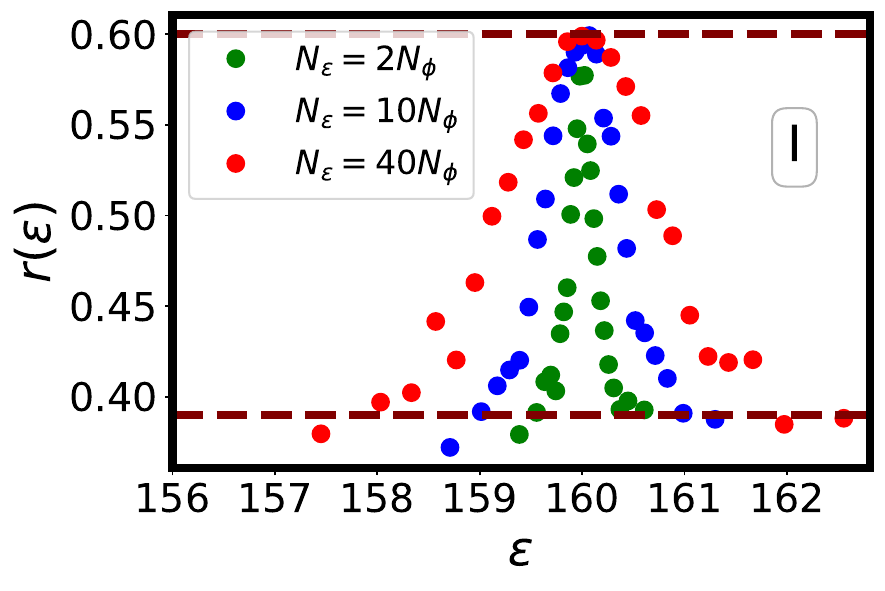}}{(a)}
\stackon{\includegraphics[width=0.33\columnwidth,height=4.4cm]{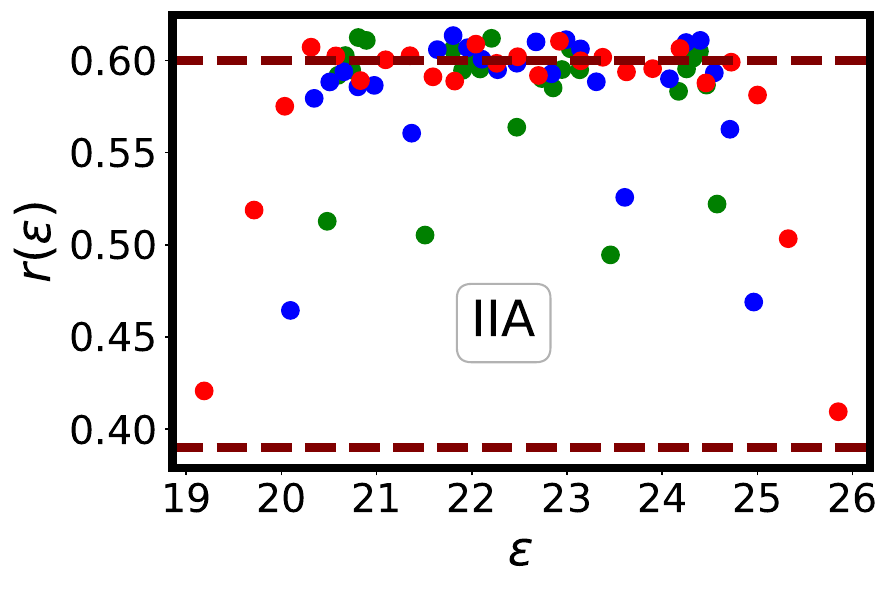}}{(b)}
\stackon{\includegraphics[width=0.33\columnwidth,height=4.4cm]{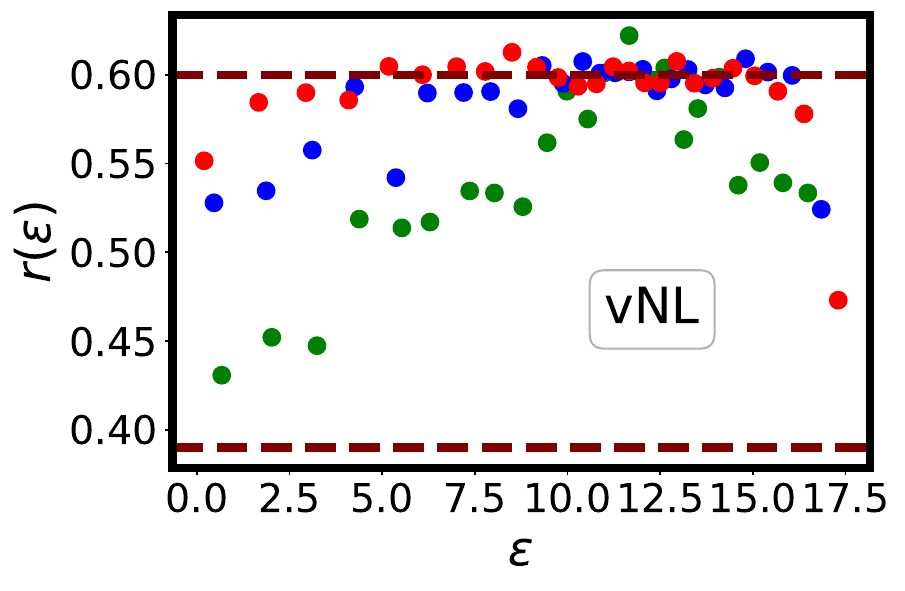}}{(c)}
\stackon{\includegraphics[width=0.33\columnwidth,height=4.4cm]{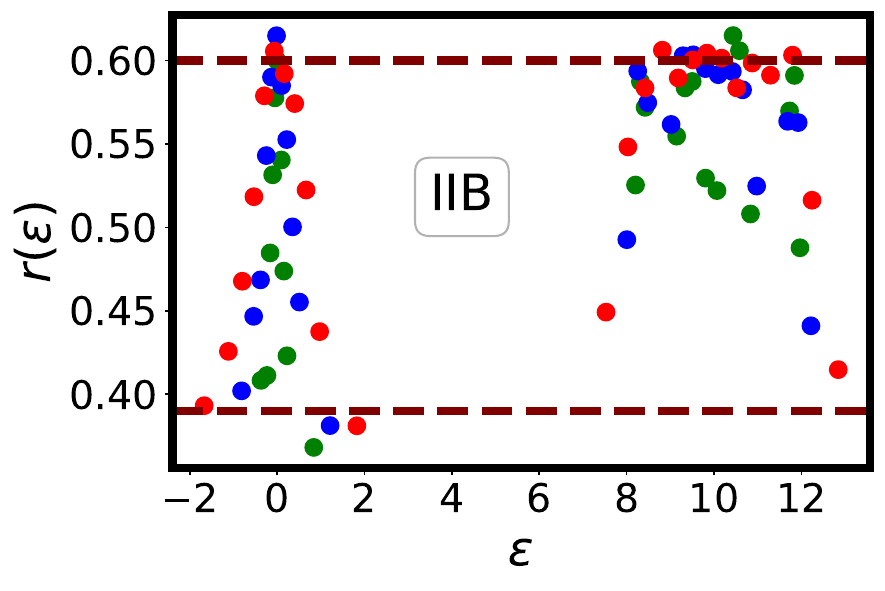}}{(d)}
\stackon{\includegraphics[width=0.33\columnwidth,height=4.4cm]{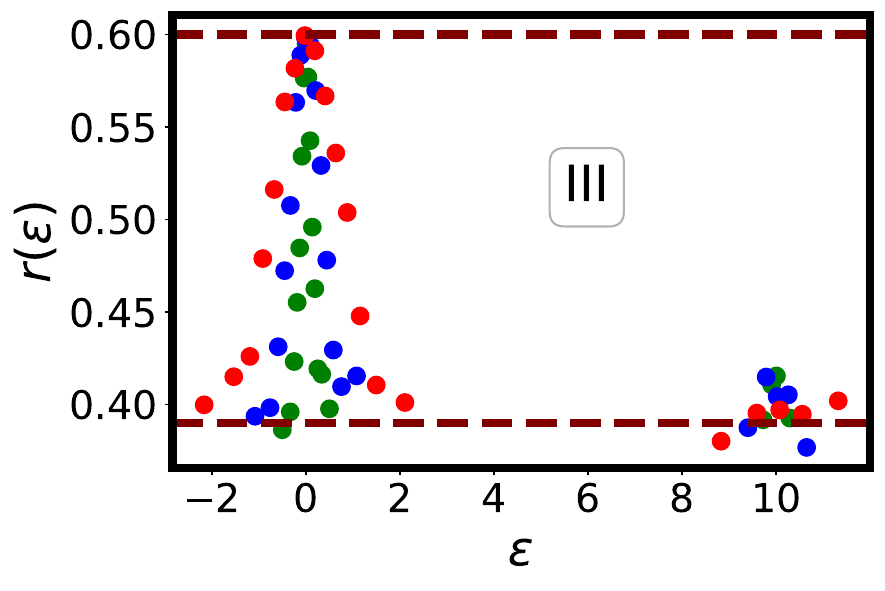}}{(e)}
\caption{ {\bf Disorder-density dependence of energy-resolved level-spacing ratio}: (a-e) Energy-resloved $r(\epsilon)$ for increasing no. of disorder potentials $N_\epsilon=2N_\phi,10N_\phi,40N_\phi$ in the regions I,IIA, at vNL point, regions IIB and III, respectively. Here $N_\phi=400$ for all the plots.}
\label{lratio_disorder}
\end{figure}

\setcounter{section}{3}
\section{~\thesection{The strong disorder case}}
\begin{figure}[h!]
\centering
\stackon{\includegraphics[width=0.246\columnwidth,height=3.7cm]{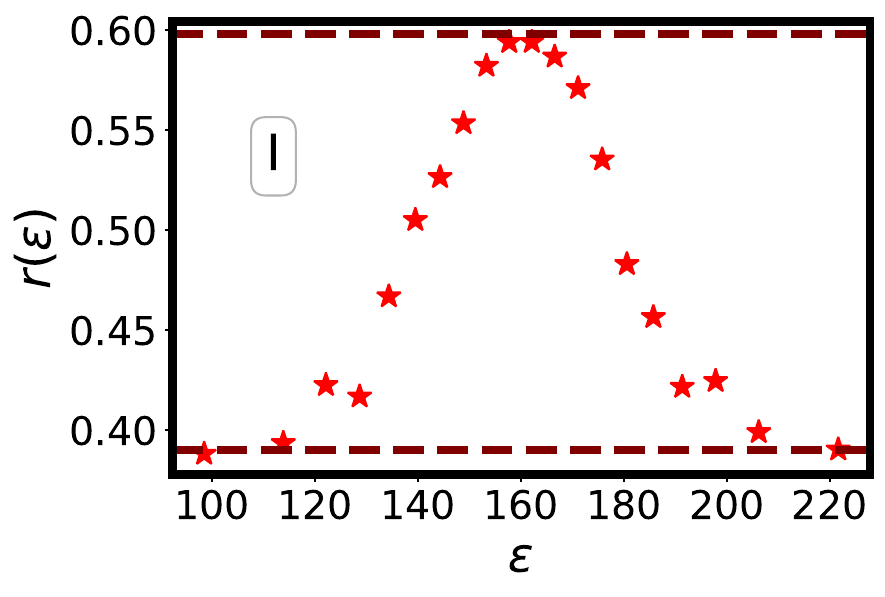}}{(a)}
\stackon{\includegraphics[width=0.246\columnwidth,height=3.7cm]{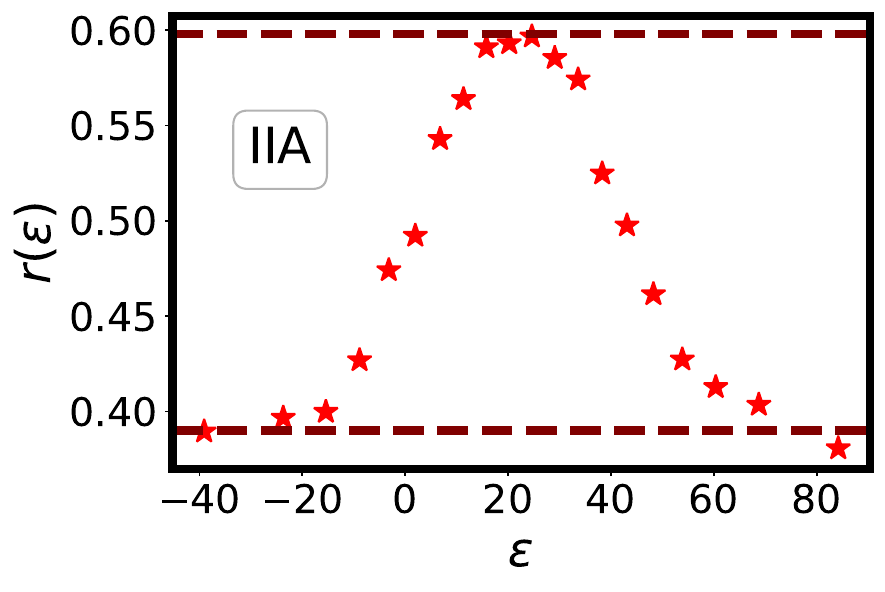}}{(b)}
\stackon{\includegraphics[width=0.246\columnwidth,height=3.7cm]{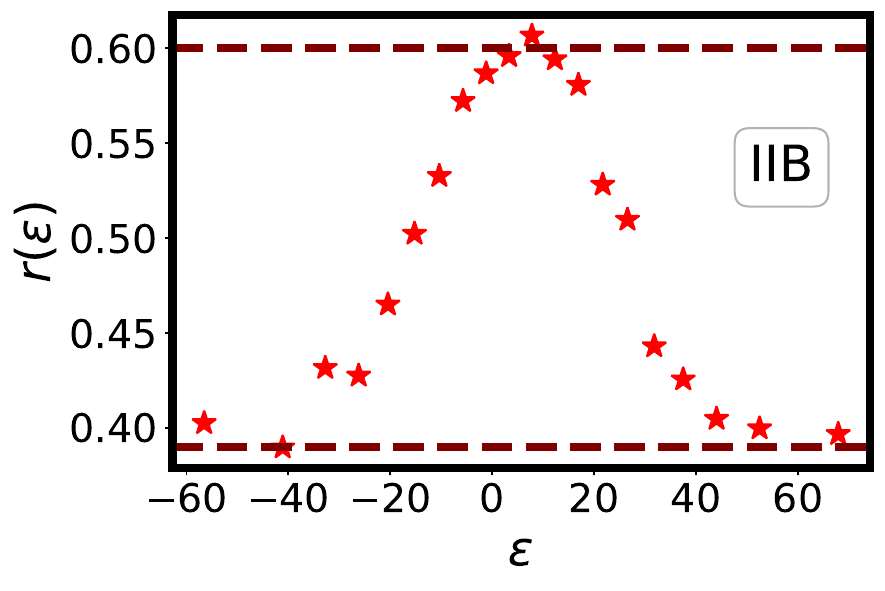}}{(c)}
\stackon{\includegraphics[width=0.246\columnwidth,height=3.7cm]{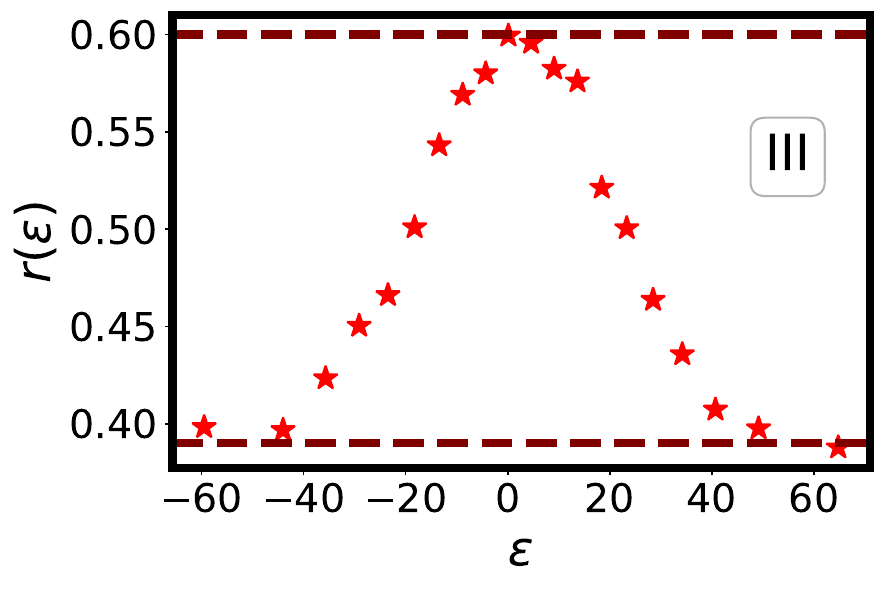}}{(d)}
\caption{ {\bf Energy-resolved levelspacing ratio in strong-disorder limit}: (a-d) Levelspacing ratio $r(\epsilon)$ as a function of single-particle energy $\epsilon$ in regions I, IIA, IIB and III, respectively in presence of strong disorder strength. Here $N_\phi=400$ and for all the plots.}
\label{lratio_strong}
\end{figure}

In Fig.~3 of main text, we have seen the effect of weak disorder on different regions depending on the value of the lattice-spacing parameter $a_s$. Here in Fig.~\ref{lratio_strong}(a-d) we show the effect of strong disorder strength in the same regions. Here we choose $W=5.0$ and $\lambda=10$ which essentially mix the non-zero energy states with zero-energy states wherever present. However, we see similar effect of strong disorder in all four regions I, IIA, IIB and III. The large fraction of delocalised states, found in regions IIA and IIB in presence of weak disorder, now shrinks to very few states in presence of strong disorder which cannot be treated perturbatively on clean system anymore. Hence, one has a tiny fraction of delocalised states responsible for a sharp integer plateau transition in Hall-resistivity experiments. This also proves the importance of DMTS for the delocalising effect in the weak disorder limit. The effect of DMTS goes away as strength of disorder approaches the strength of the lattice potential. Similar results are obtained when DMTS is broken by randomizing the positions or magnitudes of the $\delta$ potentials otherwise representing a lattice.

\setcounter{section}{4}
\section{~\thesection{Spectral form factor}}
The energy-level spacing ratio deals with the adjacent energy levels and hence it measures the short-range correlations of the energy spectrum. But there is another quantity, namely the spectral form factor (SFF) that includes both short-range and long-range correlations in the spectrum. One first defines a function $Z(\tau)=\sum\limits_{m=1}^{N} e^{-iE_m \tau}$ (Fourier transform of density of states) of fictitious time $\tau$ with  $N$ being the number of energy levels. The SFF can then be written as~\cite{haake1991quantum}
\begin{eqnarray}
K(\tau) &=& \langle Z^{*}(\tau)Z(\tau)\rangle,\nonumber\\
&=& N + \langle\sum\limits_{m\neq n} e^{-i (E_m - E_n)\tau}\rangle,
\label{sff_def}
\end{eqnarray}
where $\langle...\rangle$ here stands for an average over an ensemble, e.g. realizations of disorder in our model. In a many-body energy spectrum with GUE random matrices, $K(\tau)$ typically shows three regimes: a dip and then a linear ramp followed by a plateau~\cite{chen2018universal,liu2018spectral,ahmed2021dynamics}.  
At $\tau=0$, $K=N^2$. For very small $\tau<\tau_{D}$, the Thouless time, $\langle Z^{*}(\tau)Z(\tau)\rangle=\langle Z^{*}(\tau)\rangle\langle Z(\tau)\rangle$ due to absence of spectral correlation and $K(\tau)$ decreases showing non-universal model-specific spectral features. At $\tau=\tau_H$, the Heisenberg time, $\tau$ becomes comparable to inverse of mean levelspacing and hence the second term in  Eq.~\ref{sff_def} vanishes on average and a plateau at $K=N$ appears. For $\tau_D<\tau<\tau_H$, $K$ shows a linear ramp indicating the development of spectral correlation such that $\langle Z^{*}(\tau)Z(\tau)\rangle \neq \langle Z^{*}(\tau)\rangle\langle Z(\tau)\rangle$. This region captures universal features and does not appear in the absence of any spectral correlations.  

\begin{figure}
\centering
\stackon{\includegraphics[width=0.4\columnwidth,height=5.3cm]{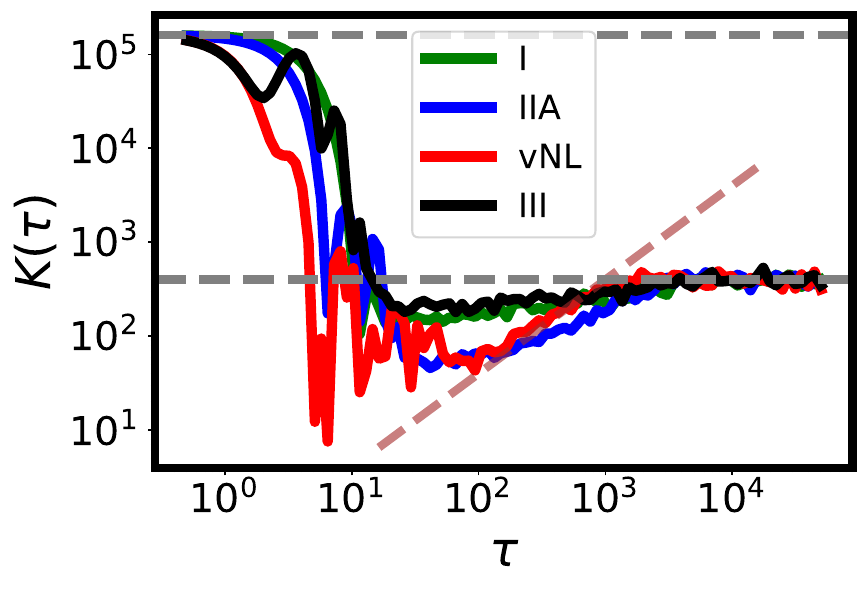}}{(a)}\hspace{0.8cm}
\stackon{\includegraphics[width=0.4\columnwidth,height=5.3cm]{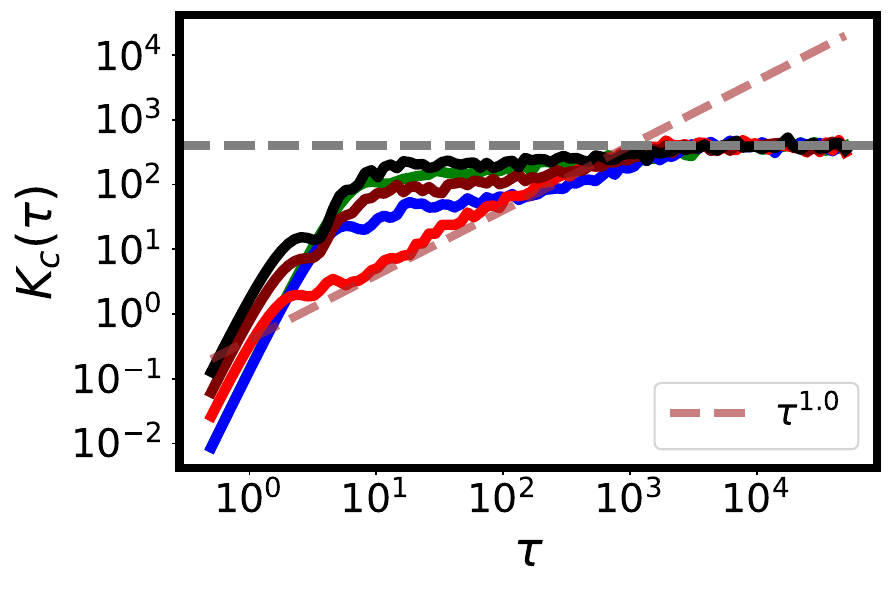}}{(b)}
\caption{ {\bf Spectral form factor}: (a) The SFF $K(\tau)$ is plotted as a function of $\tau$ for regions I, IIA, III and at the vNL limit, respectively. The upper and lower dashed horizontal lines represent $K=N_{\phi}^2$ and $K=N_\phi$, respectively.  (b) Similarly, the CSFF $K_c(\tau)$ is plotted as a function of $\tau$ for the same regions. The dashed horizontal line represent $K_c=N_\phi$. Here $W=0.2$, $\lambda=10$ and $N_\phi=400$ for all the plots. The tilted dashed line is linear in $\tau$ and provides a guide to the eyes.}
\label{sff_csff}
\end{figure}

In Fig.~\ref{sff_csff}(a) SFF is plotted for regions I, IIA, III and at vNL limit, respectively of our model which involves single particle physics in the bulk. All the plot shows a dip in the beginning and a saturation in the end. The region I and region III show a weak ramp which is not really linear. This is expected since very few delocalized states are present in the energy spectrum that leads to small amount of spectral correlations. However, in the intermediate region II the spectrum is dominated by a large fraction of delocalized states and hence one expects to see a more sharper and closer to linear ramp in this region. Expectedly we see sharper ramps for region IIA and at vNL point. In fact, for vNL the ramp is closest to the linear one since at this point the spectrum has the largest fraction of delocalized states as evinced in Fig.~\ref{sff_csff}(a). But there is also an oscillatory regime prior to the ramp, especially for the vNL. These pre-ramp oscillations are actually signatures of single-particle chaos, which has been put forth recently in the studies of quadratic SYK models~\cite{winer2020exponential,liao2020many} and SSH chains~\cite{sarkar2023spectral} which belong to symmetry-protected topological (SPT) phase. But here we find similar signatures in a two dimensional topological phase that goes beyond SPT. The non-universal spectral signatures can mask the ramp which can lead to reduced ramp size in $K(\tau)$. Hence, one typically defines the connected SFF (CSFF) which can be written as
\begin{eqnarray}
K_c(\tau) &=& \langle Z^{*}(\tau)Z(\tau) \rangle - \langle Z(\tau) \rangle \langle Z^{*}(\tau) \rangle,
\label{csff_def}
\end{eqnarray}
which is obtained by deducting the disconnected part from $K(\tau)$. At $\tau=0$, $K_c=0$ whereas for $\tau>\tau_H$ $K_c$ is expected to saturate at $N$. The CSFF is shown in Fig.~\ref{sff_csff}(b) for the same regions as in Fig.~\ref{sff_csff}(a). Broadly we see that the initial non-universal signatures including the oscillatory regime of $K(\tau)$ do not appear in $K_c(\tau)$ anymore. Instead one now finds an extended and unmasked ramp which becomes almost linear for vNL indicating that almost all the eigenstates can become delocalized at this point and the eigenvalues resembles that of GUE matrices. This is consistent with the findings from the levelspacing ratio discussed in the main text. An extensive study of SFF and CSFF in the disordered topological phases is beyond the scope of the current work and worth exploring in the future. 

\setcounter{section}{5}
\section{~\thesection{Chern bands of Chern insulators in presence of disorder}}
\begin{figure} [h]
\centering
\stackon{\includegraphics[width=0.37\columnwidth,height=5.0cm]{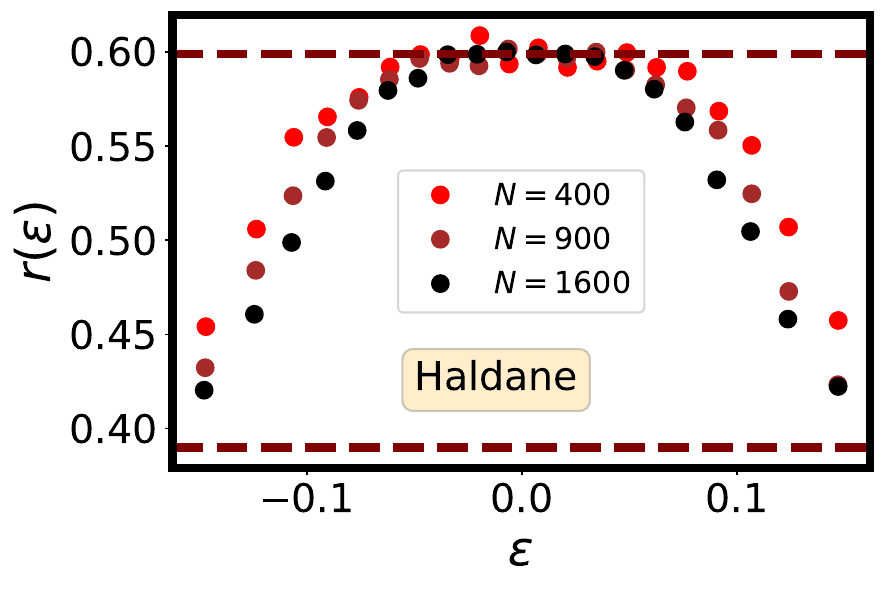}}{(a)}\hspace{0.8cm}
\stackon{\includegraphics[width=0.37\columnwidth,height=5.0cm]{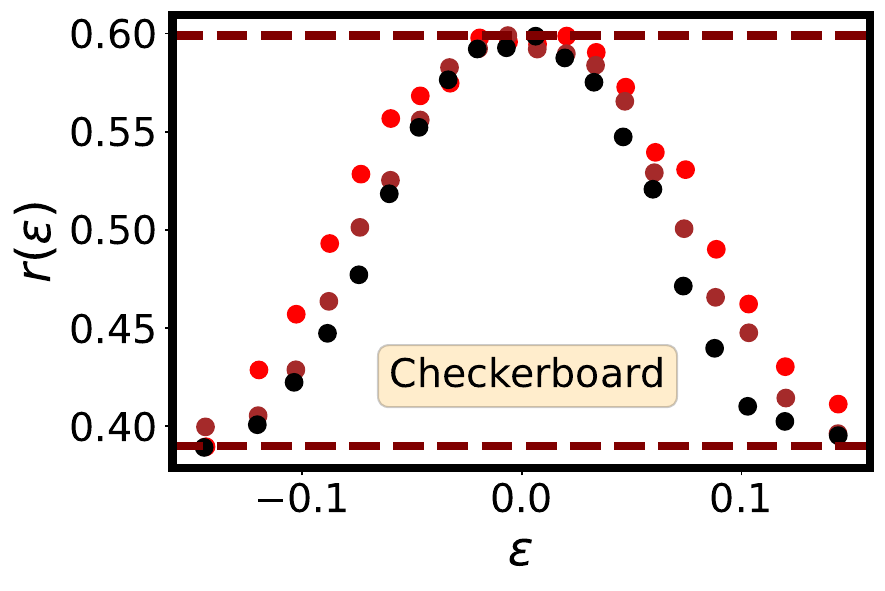}}{(b)}
\caption{ {\bf Energy-resolved level spacing ratio in disordered Chern insulators}: (a) The energy-resolved level spacing ratio $r(\epsilon)$ showing the fate of the Chern band in the Haldane model under the influence of disorder for increasing no. of unit cells $N$. (b) The same but now for the Chern band of the checkerboard lattice model. These show the appearance of single particle states with different localization properties, especially in the middle and edges of the bands, of these two models.}
\label{spacing_CI}
\end{figure}

In the main text, we have mentioned how the disordered Chern band (in the flat band limit i.e. kinetic energies are thrown away and infinite band gap is assumed) of the Haldane model~\cite{haldane1988model} can be different from a disordered Landau level obtained in IQHE. 
In fact, we have shown that the fraction of extended states obtained in the middle of the band in Haldane model is larger (Fig. 4(d) of main text) unlike the same in the Landau level where the this fraction is smaller.
We reaffirm the results for the disordered Chern band of the Haldane model in Fig.~\ref{spacing_CI}(a) through plotting the energy-resolved level spacing ratio $r(\epsilon)$ with increasing no. of unit cells $N$. 
This indicates toward the role of the non-uniform Berry curvature in the Chern band of Chern insulators unlike the Landau level where it is uniform. This further indicates that Chern bands with different distributions of Berry curvature would show different localization properties in presence of disorder. 
As an example, we consider the Chern band of the checkerboard lattice model~\cite{sun2011nearly} with comparatively less non-uniform Berry curvature, in the flat-band limit as discussed earlier. In presence of disorder, as it is shown in Fig.~\ref{spacing_CI}(b), $r(\epsilon)$ as a function of energy $\epsilon$ looks quite different from the same in the Haldane model. Unlike the disordered Chern band in the Haldane model, here we observe a much higher localization tendencies near the edges and middle of the energy spectrum, respectively as $N$ is increased. The intricate connection between the distribution of Berry curvature in Chern band and its localization properties in relation to the superlattice spacing will be discussed in details in the future.

\end{document}